\documentclass[preprint] {article}
\usepackage[utf8]{inputenc} 
\usepackage{amsmath}
\usepackage{amsfonts}
\usepackage{amssymb}
\usepackage{geometry}
\allowdisplaybreaks
\newcommand {\bp}{\begin{pmatrix}}
\newcommand {\ep}{\end{pmatrix}}
\newcommand{\be}{\begin{equation}} \newcommand{\ee}{\end{equation}}
\newcommand{\bea}{\begin{eqnarray}}\newcommand{\eea}{\end{eqnarray}}

\usepackage{geometry} 
\begin{document}
\title{Integrable Local and Non-local Vector Non-linear Schrödinger  Equation with Balanced loss and Gain}

\author{Debdeep Sinha\footnote{{\bf email:}  debdeep.sinha@saha.ac.in}}
\date{Theory Division\\
Saha Institute of Nuclear Physics\\
1/AF Bidhannagar, Calcutta - 700064, India.}
\maketitle
\begin{abstract}
The local and non-local vector Non-linear Schrödinger Equation (NLSE) with a general cubic non-linearity
are considered in presence of a linear term characterized, in general, by a non-hermitian matrix which under certain
condition incorporates balanced loss and gain and a linear coupling between the complex fields of the governing 
non-linear equations. It is shown that the systems posses a Lax 
pair and an infinite number of conserved quantities and hence integrable. Apart from the particular form of the local and 
non-local reductions, the systems are integrable when the matrix representing the linear 
term is pseudo hermitian with respect to the hermitian matrix comprising the generic cubic 
non-linearity. The inverse scattering transformation method is employed to find exact soliton 
solutions for both the local and non-local cases. The presence of the linear term
restricts the possible form of the norming constants and hence the polarization vector.
It is shown that for integrable vector NLSE with a linear term, characterized by a pseudo-hermitian 
matrix, the inverse scattering transformation selects a particular class of solutions of the corresponding vector NLSE without
the linear term and map it to the solution of the integrable vector NLSE with the linear term
via a pseudo unitary transformation, for both the local and non-local cases.

\end{abstract}

\section{Introduction}

The NLSE being an integrable and exactly solvable model has always been a focal point
 of mathematical investigation to understand nonlinear evolution equations and over the years many
methods have been developed exploring the intriguing richness of such systems. Further importance
of this model arises due to the fact that NLSE and its various generalizations are capable of describing many physical
phenomena encompassing a diverse branches of physics that include wave propagation in non-linear media \cite{ac},
 Bose-Einstein condensation (BEC) \cite{r2}, plasma physics \cite{r3}, gravity waves \cite{r4}, 
$\alpha$-helix protein dynamics \cite{r5} etc. The initial studies on NLSE was confined to homogeneous and 
autonomous systems where it describes the wave propagation in non-linear optical media. As soon as
it is realized that various generalizations of NLSE can also be integrable, its application enhances to include 
inhomogeneous and non-autonomous solitons. Such systems describe many interesting physical phenomenon 
that find applications in the study of
BEC, soliton lasers, ultrafast soliton switches, and logic gates \cite{cl, vn, r6,r8,r9,ag}. Recently, the emergence
of Parity-Time (${\cal PT}$) symmetric systems and its growing relevance in the nonlinear optics necessitate the
study of various generalizations of NLSE in the realm of ${\cal PT}$- symmetric systems as well \cite{r1, r10}. 
In case of nonlinear optics the confining potential is made to be ${\cal PT}$- symmetric and a great deal of 
investigation to find soliton solution is devoted in this direction that involves, some interesting study such as, the
${\cal PT}$- symmetry breaking as a function of system parameter and the behavior of soliton near the exceptional
points \cite{su, zh, zh1}. A note worthy advancement in the possible form of NLSE in the realm of ${\cal PT}$- symmetric 
systems is the introduction of a new type of non-local reduction for NLSE which is shown to be integrable \cite{ab} 
with an interesting feature 
of admitting both bright and dark soliton solutions with the same kind of attractive interaction. The corresponding 
vector non-local NLSE is also shown to be integrable \cite{ds}. A great deal of investigation has also been carried out 
with non-local NLSE \cite{ds2,abzm,asf,aks, akas, akmu, li, lym}. Another variety of NLSE, in case of ${\cal PT}$- symmetric systems,
is the balanced loss and gain systems with linear coupling that appears in the study of coupled and multi
coupled wave guide systems described by multi component NLSE. Recent development in this direction may be found in
the following Refs. \cite{xi, igor, yu, ro, alj}.

The purpose of the present paper is to investigate the integrable properties of local and non-local vector
NLSE with generic cubic non-linearity and with a linear term characterized, in general, by a non-hermitian matrix
which under certain condition incorporates balanced loss and gain with a linear coupling between the components of the complex fields 
corresponding to the vector nonlinear equations.
Recently, a systematic investigation of NLSE with balanced loss and gain and with linear coupling term is 
presented in Ref. \cite{pkg}. It is shown that the system can be solved exactly via a non-unitary 
transformation. In particular, a vector NLSE with balanced  loss-gain, linear coupling and a general form of  
cubic non-linearity is considered  \cite {pkg}. A non-unitary transformation is used to map the system to the 
same equation without the loss-gain and linear coupling terms with a modified time-modulated non-linear 
interaction. Further, the non-linear term remains invariant for the special case of pseudo unitary transformation.
The non-unitary transformation considered in this case may be viewed as a gauge transformation involving complex
scalar gauge potential. However, since the original system and mapped system are connected via non-unitary 
transformation the observables have different time dependence and the two systems are not gauge equivalent.
Further investigation is also made to construct exactly solvable models for the case of time dependent loss and 
gain and space time modulated non-linear interaction \cite{sg}. It should be mentioned here that the NLSE with 
balanced loss and gain arises in the study of non-linear optics and admits bright as well as dark soliton solutions\cite{yu,ro, rr10},
rouge waves \cite{rw}, exceptional points \cite{ex1, ex2} and breathers solitons \cite{ex1}. Some of the models involving balanced 
loss and gain are also investigated from the viewpoint of exactly solvable models \cite{ro, igor1}. 
Numerical investigation has also been made to exhibit the efficiency in the stabilization of solitons and 
soliton switching in case of NLSE with time dependent balanced loss and gain.

It should be mentioned that the study of the integrable properties of vector NLSE incorporating balanced loss and gain and 
linear coupling is rather limited, albeit, some exactly solvable models are obtained. Therefore, it is affirmative to study the
integrable properties of such systems. In the present work investigation is made to study the integrable properties of vector NLSE 
with a linear term characterized, in general, by a non-hermitian matrix. In particular, a system of vector NLSE with general 
form of cubic interaction that includes self-phase and cross-phase modulation along with four wave mixing is considered 
with a linear term characterized by a non-hermitian matrix. A particular choice of this matrix generates the balanced loss 
and gain system with a linear coupling between the complex fields of the governing non-linear equation. 
Further, a non-local vector NLSE is also considered in presence of a linear term. This may be viewed as a two fold
generalization of the model considered in Ref. \cite{ds}. In particular, it generalizes the cubic interaction to include the
self-phase and cross-phase modulation along with four wave mixing. Further, the integrability of this model is investigated 
in presence of a linear term characterized by a non-hermitian matrix which under certain condition represents balanced 
loss and gain with a linear coupling between the complex fields of the governing non-linear equation. 
The Lax-pair of the systems are constructed for both local and non-local cases and an 
infinite number of conserved quantities are found which confirm the integrability of the systems. Apart from the particular
form of the reductions, the systems are integrable when a specific condition is satisfied. In particular, the systems are integrable 
only when the matrix representing the linear term is pseudo hermitian with respect to the hermitian matrix governing 
the cubic non-linearity of the systems. This is an important result 
in the realm of integrability of vector NLSE in presence of a generic linear term. It is interesting to note that the condition of 
pseudo hemitisity on the matrix representing the generic linear term of the vector NLSE, also appears as a condition 
for the system to be mapped to the same equation without the loss-gain and linear coupling terms 
while the non-linear term remains invariant \cite{pkg}. The inverse scattering transformation method is employed to find exact soliton 
solutions for the local and non-local vector NLSE in presence of a linear term characterized, in general, by a non-hermitian matrix. 
An important finding of the present investigation is that the the presence of the linear term restricts the possible form of the
norming constants and hence the polarization vector. It is shown that for integrable vector NLSE with a linear term characterized, 
by a pseudo hermitian matrix, the inverse scattering transformation selects a particular class of solutions of the corresponding vector
NLSE without the linear term and map it to the solution of the integrable vector NLSE with the linear term via a pseudo unitary 
transformation, for both the local and non-local cases. It should be mentioned that the mapping of the solution of a particular 
system to that of an integrable one does not necessarily imply the integrability
of the system under consideration. Two integrable systems are said to be gauge equivalent if the corresponding Lax-pairs are related by a 
gauge transformation. For the present case this gauge transformation to the integrable vector NLSE is not apparent. 
Further, the transformation performed in Ref. \cite{pkg} to map the system to an integrable one is not unitary. Therefore, the study 
of integrability of the local and non-local vector NLSE with generic cubic non-linearity in presence of a linear term characterized, in general, by a 
non-hermitian matrix, is an important addition to the existing list of integrable systems.

The plan of the paper is as follows. In the next section the models are introduced and the Lax-pair is constructed
for the local and non-local vector NLSE in presence of a generic linear term. In section-3, 
general results concerning the inverse scattering transformation are presented and the time evolution of the scattering data
is considered. The one soliton solution and the conserved quantities for the local and non-local cases are respectively presented
in section-4 and in section-5. In the last section, we make a summary and discuss the results. 

\section{Introduction to the models and the construction of the Lax-pair}

In this section, we construct the Lax-pair for the local and non-local vector NLSE with a generic form of cubic non-linearity in
presence of a linear term. A natural generalization of the vector NLSE  may be expressed in the following
manner

\bea
i {\bf Q}_t- {\bf Q}_{xx}-{\bf AQ}+ 2{\bf Q}{\bf RQ}=0
\label{eq1}
\eea
where ${\bf Q}_t= \frac{d{\bf Q}}{dt}$, ${\bf Q}_{xx}= \frac{d^2{\bf Q}}{dx^2}$ 
and ${\bf Q}= (q_1(x,t), q_2(x,t))^{T}$, with $T$ denoting the transpose of a 
matrix, is a two component column vector with $q_1(x,t)$ and $q_2(x,t)$ being
two complex fields with real arguments $x$ and $t$. The two component row 
vector has the form ${\bf R}=(r_1(x,t), r_2(x,t))$ where $r_1(x,t)$ and $r_2(x,t)$ are two
complex fields with real arguments $x$ and $t$. The specific nature of ${\bf R}$ 
determines the form of the nonlinear interaction. For example, we shall be considering 
two specific choices for ${\bf R}$ corresponding to local and non-local reduction 
respectively, involving the cross-phase and self-phase modulation along with four-wave mixing.
The matrix ${\bf A}$ is a constant matrix that determines the linear interaction
between the components of complex vector field ${\bf Q}$, with the diagonal terms are
responsible for the self-coupling and the off-diagonal terms are responsible for the cross coupling. 
The interesting case of balanced loss and gain occurs when the diagonal elements of {\bf A} become complex
with the imaginary parts of the diagonal elements add to zero.
For ${\bf A}=0$, the system described by Eq. (\ref{eq1}) reduces to the ordinary vector NLSE  of specific form
depending on choice of the reduction for ${\bf R}$.
The local vector NLSE with a linear term arises for the reduction:

\bea
{\bf R}= \mp({\bf GQ})^{\dagger}
\label{re1}
\eea
where ${\bf G}$ is a non-singular hermitian matrix of constant arguments and $'\dagger'$ denotes the transpose
 with complex conjugation. Eq. (\ref{eq1}) with the reduction of Eq. (\ref{re1}) gives the local vector 
NLSE with a linear term and have the following form:

\bea
i {\bf Q}_t- {\bf Q}_{xx}-{\bf AQ}\mp2\left({\bf Q}^{\dagger}{\bf GQ}\right){\bf Q}=0.
\label{lv}
\eea
The hermitian matrix ${\bf G}$ includes generic cubic interaction with the diagonal terms
involving the cross-phase and self-phase modulation while the off-diagonal terms are denoting the four-wave mixing.
The integrable properties of the system for the special case of ${\bf A}=0$, i.e, the vector NLSE with cross-phase and 
self-phase modulation and four-wave mixing without the linear term has been considered in Ref.\cite{wang}. 
It may be mentioned here, that the integrability of vector NLSE without the four-wave mixing has been considered earlier
for the case when the self-phase coefficient is equal to the cross-phase coefficient \cite{ac, rl,tt} and also when the 
self-phase coefficient has the opposite sigh to that of the cross-phase coefficient \cite{sc,av}.
The recent applications of vector NLSE includes the case of coupled ${\cal PT}$-symmetric optical wave guide systems
\cite{xi,igor,yu}, the case of multi-core wave guide structures \cite{ro}, Bloch-wave packets in a periodic system \cite{bl},
spinor Bose–Einstein condensates \cite{becs,becs1} etc.. The minus and plus sign before the non-linear 
interaction term in Eq. (\ref{lv}) respectively correspond to attractive and repulsive type of interactions. In Ref. \cite{pkg}
it is shown that the model (\ref{lv}) is exactly solvable for autonomous as well as non autonomous case when the matrix
${\bf G}$ includes specific time dependence. The model is also exactly solvable for the specific time dependence of the coupling
parameters and space-time modulated nonlinear interaction term \cite{sg}. In this paper, the integrable properties of the
autonomous system (\ref{lv}) is discussed. The integrable properties of the non-autonomous version is considered in Ref. \cite{ds1}.

 The non-local vector NLSE with a linear term arises for the reduction:

\bea
{\bf R}= \mp({\bf GQ})^{P}
\label{re2}
\eea
where $P$ denotes the transpose with complex conjugation plus the parity transformation of
 the spatial argument. In particular, ${\bf Q}^{P}= (q^*_1(-x,t), q^*_2(-x,t))$ with $'*'$ denotes the
 complex conjugation. Eq. (\ref{eq1}) with the reduction of Eq. (\ref{re2}) gives the non-local vector 
NLSE with a linear term of the following form:

\bea
i {\bf Q}_t- {\bf Q}_{xx}-{\bf AQ}\mp2\left({\bf Q}^{P}{\bf GQ}\right){\bf Q}=0.
\label{nlv}
\eea
It should be noted that in contrast to the vector NLSE, here the argument in the potential term of the complex
field is $-x$ instead of $x$. This introduces a non-local nature in the potential. The system corresponding to the 
particular case of ${\bf A}=0$ and ${\bf G}= {\bf I}$, where ${\bf I}$ denotes a $2\times 2$ unit matrix, has been 
considered in Ref. \cite{ds} which is shown to be integrable and soliton solution is obtained via inverse scattering 
transformation. Eq. (\ref{nlv}) may be considered as a twofold generalization of the model considered 
earlier in Ref. \cite{ds}. One, in incorporating cross-phase and self-phase modulation and four-wave mixing term 
in the nonlinear interaction and other in considering the linear interaction between the components of the components of the
complex field ${\bf Q}$. Here, also the minus and plus sing before the potential term in Eq. (\ref{nlv}), respectively 
correspond to the attractive and repulsive type of interaction.

Eqs. (\ref{lv}) and  (\ref{nlv}) yield a Lax-pair that reduces the non-linear equations into a pair of linear equations, one of 
which gives the scattering problem and other gives the time evolution of the scattering data, and may be represented 
as

\bea
v_{x}&=& {\bf X}v,\ \ {\bf X}=
\bp
-ik {\bf I_{2}}& {\bf Q}\\
 {\bf R} & ik  {\bf I_{1}}  \\
\label{sc}
\ep
\label{vx}
,\\
v_{t}&=& {\bf T}v, \ \ {\bf T}=
\bp
2ik^2{\bf I_2}+i {\bf Q} {\bf R}-i{\bf A}& -2k {\bf Q}-i {\bf Q}_{x} \\
-2k {\bf R}+i {\bf R}_{x} &  -2ik^2 {\bf I_1}-i {\bf R} {\bf Q} \\
\ep 
\label{vt}.
\eea
Eq. (\ref{vx}) is an eigenvalue equation with a constant parameter $k$ 
and Eq. (\ref{vt}) gives the time evolution of the corresponding fields. 
In this case $v(x,t)$ is a three component column vector and it is assumed that 
the fields $\{q_1(x,t),q_2(x,t)\}$ and $\{r_1(x,t),r_2(x,t)\}$ vanish rapidly as $|x|$ 
$\rightarrow$ $ \infty$, and ${\bf I}_{n}$ denotes $n\times n$ square matrices .
The important result is that the compatibility condition $v_{xt}$ $=$ $v_{tx}$ between 
Eqs. (\ref{vx}) and (\ref{vt}) give rise to the local and non-local vector NLSE with the linear 
 term of the form of Eq. (\ref{lv}) and Eq. (\ref{nlv}) under the symmetry reduction
of Eq. (\ref{re1}) and Eq. (\ref{re2}) respectively, only when the following condition is satisfied:

\bea
{\bf A}^{\dagger}= {\bf GAG^{-1}},
\label{con1}
\eea
i.e, the matrix ${\bf A}$ is ${\bf G}$-pseudo-hermitian \cite{pseh}. This is a note worthy condition as an integrable criteria for
a class of NLSE with a linear term. So far the matrix ${\bf A}$ and ${\bf G}$, in Eqs. 
(\ref{lv}) and (\ref{nlv}), which appear to be independent, now constrained by the condition of Eq. (\ref{con1}) in order 
that the systems are integrable. Some observations are in order:\\

\noindent i) ${\bf A}$ is hermitian: In this case condition (\ref{con1}) implies that the systems are integrable only when
the matrix ${\bf A}$ and ${\bf G}$ commute, i.e, $[{\bf A}, {\bf G}]=0$. In that case there will be no loss and gain.\\

\noindent ii) ${\bf G}$ is a unit matrix: In this case ${\bf A}$ becomes hermitian and again there will be no loss and gain.\\

\noindent iii) The interesting case arises when the matrix ${\bf A}$ is non-hermitian. Further, the physically motivated case
of balanced loss and gain occurs when the diagonal elements of ${\bf A}$ become complex with the imaginary parts 
of the diagonal elements add to zero. In particular, the matrix ${\bf A}$ may be decomposed as a sum of two hermitian 
matrices ${\bf B, C}$, i.e, ${\bf A}={\bf B}+i {\bf C}$ with ${\bf C}$ being a traceless diagonal matrix representing the strength
of loss/gain. 
It should be noted that the system is still integrable if $\text{Tr}(C)\ne 0$ as far as condition (\ref{con1}) is satisfied. However,
unbalanced loss and gain leads to unstable solution and physically motivated cases arise when loss and gain are equally
balanced \cite{pkg}. The condition (\ref{con1}) also appears in Ref. 
\cite{pkg} as a condition for the system to be mapped to the same equation without the loss-gain and linear coupling terms 
while the non-linear term remains invariant and exact solutions are obtained. For the rest of the discussions, we will consider 
the case for which ${\bf A}$ is ${\bf G}$-pseudo-hermitian.

It may be noted that two integrable systems having the Lax-pairs $({\bf X}, {\bf T})$ 
and $({\bf \tilde{X}}, {\bf \tilde{T}})$ and admit the linear equations of the form $v_x={\bf X}v, v_t={\bf T}v$
and $u_x={\bf \tilde{X}}u, u_t={\bf \tilde{T}}u$ respectively, are said to be gauge equivalent, if the Lax-pairs are related by a 
gauge transformation ${\bf g}$,  in the following manner:

\bea
{\bf X}={\bf g}{\bf \tilde{X}}{\bf g}^{-1}+{\bf g}_x{\bf g}^{-1}, \ \ \ \  {\bf T}={\bf g}{\bf \tilde{T}}{\bf g}^{-1}+{\bf g}_t{\bf g}^{-1}
\label{gt}
\eea
where $v={\bf g}u$.
For example, $({\bf \tilde{X}}, {\bf \tilde{T}})$ may be chosen to represent the Lax-pair of the vector NLSE without the 
linear term, i.e, ${\bf A}=0$, where $({\bf X}, {\bf T})$ is given by Eqs. (\ref{vx}) and (\ref{vt}). In this case $({\bf X}, {\bf T})$ 
and $({\bf \tilde{X}}, {\bf \tilde{T}})$ are related by the following expressions:

\bea
{\bf X}={\bf \tilde{X}}, \ \ \   {\bf T}={\bf \tilde{T}}+{\bf \tilde{S}},\ \ \ {\bf \tilde{S}}=\bp -i{\bf A} & {\bf 0}\\ {\bf 0} & 0\ep
\label{gxt}
\eea
where ${\bf \tilde{S}}$ is a $3\times 3$ matrix. In this case, Eqs. (\ref{gt}) and (\ref{gxt}), yield the following set of 
equations for the gauge transformation matrix ${\bf g}$:

\bea
{\bf g}_x=[{\bf \tilde{X}}, {\bf g}],\ \ \ \ {\bf g}_t=[{\bf \tilde{T}}, {\bf g}]+ {\bf \tilde{S}}{\bf g}.
\label{g}
\eea
However, the solution of (\ref{g}) is not apparent. It is interesting to note that the solution of the nonlinear equation represented by
the Lax-pair $({\bf \tilde{X}}, {\bf \tilde{T}})$ are related to the solution of the nonlinear equation represented by the 
Lax-pair $({\bf X}, {\bf T})$ via a pseudo unitary transformation depending on the specific form of ${\bf A}$ \cite{pkg}.

\section{The inverse scattering transformation}

In this section, we present the general results concerning the inverse scattering transformation and obtain the
time evolution of the scattering data.
In order to solve the scattering problem of Eq. (\ref{vx}), we define the following Jost functions satisfying the
constant boundary conditions at $|x| \rightarrow \infty$:
\bea
M(x,t)=exp(ikx)\phi(x,k),\            \bar{M}(x,k)=exp(-ikx)\bar{\phi}(x,k),\nonumber\\
N(x,t)=exp(-ikx)\psi(x,k),\            \bar{N}(x,k)=exp(ikx)\bar{\psi}(x,k),
\label{jf}
\eea 
where $\phi, \bar {\psi}$ and hence M and $\bar{N}$ are $3 \times 2$ and  $\bar{\phi}, \psi $
and hence $\bar{M}$ and $N$ are $3\times 1$ matrices. The functions $(\phi, \bar{\phi})$ are
the solutions of the scattering problem of Eq. (\ref{vx}) satisfying the boundary conditions as 
$x\rightarrow -\infty$ and $(\psi, \bar{\psi})$ are the solutions satisfying the boundary conditions 
as $x\rightarrow \infty$. The asymptotic form of the functions $(\phi, \bar{\phi})$ and $(\psi, \bar{\psi})$ 
may be obtained by using the vanishing boundary condition ${\bf Q}, {\bf R} \rightarrow 0$ as 
$|x| \rightarrow \infty$ in Eq. (\ref{vx}), with the following results:

\bea
\phi(x,k)\sim
\bp
{\bf I}_2 \\
0
\ep
e^{(-ikx)},\ \ \ \ 
\bar{\phi}(x,k)\sim
\bp
0 \\
{\bf I}_1
\ep
e^{(ikx)}\ \  \text{as} \ \ \ x \rightarrow -\infty \nonumber\\
\psi(x,k)\sim
\bp
0\\
{\bf I}_1
\ep
e^{(ikx)},\ \ \ \ 
\bar{\psi}(x,k)\sim
\bp
{\bf I}_2\\
0
\ep
e^{(-ikx)}\ \  \text{as} \ \ \ x \rightarrow \infty.
\label{bc}
\eea
By computing the Wronskian and using the asymptotic from of Eqs. (\ref{bc}) as $x \rightarrow \mp \infty$, 
it can be shown that the functions $(\phi, \bar{\phi})$ and $(\psi, \bar{\psi})$ are two sets of linearly independent
solutions of  the scattering problem satisfying the boundary  conditions at $x \rightarrow \mp \infty$ respectively. 
Further, it can be shown that if  ${\bf Q}, {\bf R} \in L^1({\mathbb R})$, then the functions $M, N$ are analytic in the 
upper half whereas $\bar{M}, \bar {N}$ are analytic in the lower half of the complex k-plane \cite{aab}. An integral 
expression for the Jost functions \cite{aab} with the constant boundary conditions induced by Eqs. (\ref{bc}) can be obtained 
by using the expressions of Eqs. (\ref{jf}) in the scattering problem of Eq. (\ref{vx}). Since $\{\phi,\bar\phi\}$
and $\{\psi,\bar\psi\}$ are the two sets of linearly independent solutions of the scattering problem one can express 
one set of solutions in terms of the linear combinations of the other. Thus we can write

\bea
\bp
\phi(x,k)&
\bar{\phi}(x,k)
\ep
=
\bp
\bar{\psi}(x,k)&
\psi(x,k)
\ep
\bp
{\bf a} & {\bf \bar{b}}\\
{\bf b} & {\bf \bar{a}}
\ep
\label{l}
\eea

\noindent where ${\bf{b}}(k)$ is $1 \times 2$, ${\bf{a}}(k)$ is $2\times2$, $\bar{\bf{a}}(k)$ is
 $1\times1$ and $\bar{\bf{b}}(k)$ is $2\times1$ matrix.

The inverse scattering problem deals with the reconstruction of the potentials by using the RH
(Riemann-Hilbert) approach. Using Eq. (\ref{l}) one can obtain, by exploiting the 
analytic properties of the Jost functions, the following
equations for ${\bf N}(x,k)$ and ${\bf \bar{N}}(x,k)$ 

\bea
{\bf N}(x,k)=
\bp
{\bf 0}\\
{\bf I_1}
\ep
+ \sum^{\bar{J}}_{j=1} \frac{e^{-2i{\bar{k}_j}x}{\bf \bar{N}_j}(x){\bf\bar{C}_j}}{(k-{\bar{k}_j})},
\ \ \ \ 
\bar{{\bf N}}(x,k)=
\bp
{\bf I_2}\\
{\bf 0}
\ep
+ \sum^{J}_{j=1}\frac{e^{2ik_jx}{\bf N_j}(x){\bf C_j}}{(k-k_j)} 
\label{e6}
\eea 

\noindent where ${\bf N_j}(x)={\bf N}(x,k_j)$, ${\bf \bar{N}_j}(x)={\bf \bar{N}}(x,\bar{k}_j)$ and $k_j$ and $\bar{k}_j$ 
are the proper eigenvalues of the scattering problem in upper and lower half of the complex $k$-plane respectively. 
The proper eigenvalues of the scattering problem is defined as the complex
value of $k$ for which the scattering problem of Eq. (\ref{vx}) admits a bounded solution that decays as $x \rightarrow 
\pm \infty$. The norming constants are given by the expressions ${\bf C}_j=\frac{{\bf b}(k_j){\bf \alpha}(k_j)}{(\det{{\bf a})'}(k_j)}$, 
${\bf \bar{C}}_j=\frac{{\bf\bar{b}(\bar{k}_j) \bar{\alpha}(\bar{k}_j)}}{(\det {\bar{{\bf a}})}'{(\bar{k}_j)}}$, where $'$ denotes
the derivative with respect to $k$ and $\alpha$ and $\bar{\alpha}$ are respectively the transpose of cofactor of $\bf a$ and $\bf \bar{a}$.
Further, in the present case, it is assumed that ${\bf ba^{-1}}={\bf \bar{b}\bar{a}^{-1}}=0$. 
In order to obtain the solution for ${\bf Q}, {\bf R}$, the asymptotic expressions for $N, \bar{N}$ of Eq. (\ref{e6}) are 
compared to that of the asymptotic expressions of the same functions obtained by solving the scattering problem for large $k$. 
In particular, the following results are obtained:

\bea
{\bf Q}=2i\sum^{\bar{J}}_{j=1} e^{-2i\bar{k}_j x}
\bp
\bar{N}_{11}\bar{C}_1+ \bar{N}_{12}\bar{C}_2\\
\bar{N}_{21}\bar{C}_1+ \bar{N}_{22}\bar{C}_2
\ep
,\ \ {\bf R}= -2i \sum^{{J}}_{j=1} e^{2i {k}_j x} {\bf C}N_3
\label{qr}
\eea
where ${\bar{N}_{ij}}, i,j=1,2$ and $N_3$ are respectively the components of the respective matrices.

In order to obtain the time dependence of the scattering data, the following time dependent functions are defined 

\bea
\Phi &=& \phi(x,t) e^{(2ik^2 {\bf {I}}_{2} - i{\bf A)}t},\   \ \bar{\Phi}=e^{-2ik^2 t}\bar{\phi}(x,t)\nonumber\\
\Psi&=&e^{-2ik^2t}\psi(x,t),\   \ \bar{\Psi}=\bar{\psi}(x,t) e^{(2ik^2 {\bf I}_{2}-i{\bf A})t}
\label{tf}
\eea
as solutions of the time evolution Eq. (\ref{vt}). Therefore, the time dependence of the sets 
 $\{\phi,\bar\phi\}$ and $\{\psi,\bar\psi\}$ can now be obtained by substituting Eq. (\ref{tf}) in Eq. (\ref{vt}),
with the following results

\bea
\phi_t &=& T\phi- \phi \left(2ik^2 {\bf I_{2}} - i{\bf A}\right), \ \ \ \  \bar{\phi}_t =T\bar{\phi}+ 2ik^2 \bar{\phi}\nonumber\\
\psi_t&=&T\psi+2ik^2\psi, \ \ \ \  \bar{\psi}_t=T\bar{\psi}- \bar{\psi} \left(2ik^2 {\bf I_{2}}-i {\bf A}\right).
\label{tev}
\eea
Taking the time derivative of both side of Eq. (\ref{l}), using the results of (\ref{tev}) and employing the boundary
conditions that $\{\psi,\bar\psi\}$ satisfy as $x \rightarrow \infty$, the following time dependence of the 
scattering constants are obtained

\bea
{\bf a}_t&=& -i[{\bf A},  {\bf a}], \ \ \ \ \ \  {\bf{b}}_t= {{\bf{b}}(-4ik^2{\bf I_2}+i{\bf A})}\nonumber\\
{\bf \bar{a}}_t&=&0,\ \ \ \ \ \  \ \ \ \ \ {\bf{\bar{b}}}_t= (4ik^2{\bf I_2}-i{\bf A}) {\bf{\bar{b}}}.
\label{tsc}
\eea
It is convenient to chose ${\bf a}$ as a constant matrix commutating with ${\bf A}$. This is an important condition in the
study the integrability of vector NLSE incorporation generic form of linear interaction in presence of a linear term
characterized by the matrix ${\bf A}$, and may be expressed as

\bea
[{\bf A},  {\bf a}]=0.
\label{ts}
\eea
With this choice, matrix ${\bf a}$, ${\bf \bar{a}}$ become constant. For vector NLSE without the linear term, 
${\bf A}=0$, and condition (\ref{ts}) is automatically satisfied. However, for vector NLSE with a 
 linear term, condition (\ref{ts}) has important consequences in fixing the possible form of the norming constants
as will be seen later. The solutions for ${\bf{b}}$ and ${\bf \bar{b}}$ can readily
be obtained from Eq. (\ref{tsc}) with the results, ${\bf b}={\bf{b}}(0)\exp{[(-4ik^2{\bf I_2}+i{\bf A})t]}$,
${\bf \bar{b}}=\exp{[(4ik^2{\bf I_2}-i{\bf A})t]}{\bf{\bar{b}}}(0)$, where ${\bf{b}}(0)$, ${\bf{\bar{b}}}(0)$ are
constant vectors of appropriate dimensions. The time dependence of the norming constants are given by

\bea
{\bf C}_j= {\bf C}_j(0) e^{(-4ik_j^2{\bf I}_2+i{\bf A})},\ \ \ \  {\bf \bar{C}_j}= e^{(4i\bar{k}_j^2{\bf I}_2-i{\bf A})}{\bf\bar{C}_j}(0),
\label{nc}
\eea
where ${\bf C}_j(0)$ and ${\bf\bar{C}_j}(0)$ are the integration constants. For vector NLSE without the linear term, the constants
${\bf C}_j(0)$ and ${\bf\bar{C}_j}(0)$ remain arbitrary. However, for vector NLSE involving a linear term, condition
(\ref{ts}) fixes the constants ${\bf C}_j(0)$ and ${\bf\bar{C}_j}(0)$ as will be seen during the construction of the conserved quantities.

\section{ One soliton solution for the reduction ${\bf R}= -({\bf GQ})^{\dagger}$}

In this section, the one soliton solution for the reduction ${\bf R}= -({\bf GQ})^{\dagger}$ is presented and
the conserved quantities are obtained.
The reduction ${\bf R}= -({\bf GQ})^{\dagger}$ induces a symmetry in the scattering data with
the results $\bar{k}=k^*$ and ${\bf C}=-({\bf G\bar{C}})^{\dagger}$.
The one soliton solution is obtained by taking $J=\bar{J}=1$ and closing the system by putting
$k=k_1$ and $k=\bar{k}_1$ in Eq. (\ref{e6}). The expressions of 
${\bf N}$ and ${\bf \bar{N}}$, thus obtained, can now be used in Eq. (\ref{qr}) together with the time 
dependence of the norming constants as given by  Eq. (\ref{nc}) to obtain the following expressions for the 
potentials (omitting the subscript for conveniences):

\bea
{\bf Q}= \frac{2ie^{-2ik^*x}e^{\{4i(k^*)^2{\bf I}_2-i{\bf A}\}t}{\bf\bar{C}}(0)}{1- \frac{e^{2i(k-k^*)x}e^{-4i\{k^2-(k^*)^2\}t}||{\bf \bar{C}}||^2}{(k-k^*)^2}},\ \ \ \
{\bf R}= \frac{-2ie^{2ikx}{\bf C}(0)e^{(-4ik^2{\bf I}_2+i{\bf A})t}}{1- \frac{e^{2i(k-k^*)x}e^{-4i\{k^2-(k^*)^2\}t}||{\bf \bar{C}}||^2}{(k-k^*)^2}}
\eea
where ${||{\bf \bar{C}}||^2 =\bf \bar{C}}^{\dagger}(0){\bf G}{\bf\bar{C}}(0)$.  It should be noted that $||{\bf \bar{C}}||^2$ is not
necessarily positive definite for any ${\bf G}$ as in the case when ${\bf G}$ is a unit matrix. In case $||{\bf \bar{C}}||^2<0$, the 
solution for ${\bf Q}$ encounters singularity. Therefore, we take
$ ||{\bf \bar{C}}|| ={2\eta}e^{2\delta}>0$. The expression for ${\bf Q}$ can be further 
simplified by taking $k=\xi+i\eta, \xi,\eta \in  \mathbb{R}\ \  \text{and} \ \  \eta > 0$, in this case
the solution for ${\bf Q}$ reduces to the following form

\bea
{\bf Q}= e^{-i {\bf A}t} 2 \eta e^{-i\{2\xi x-4(\xi^2-\eta^2)t-\frac{\pi}{2}\}}\text{Sech}{(2\eta x-8\xi \eta t-2\delta)} {\bf P},
\eea
where ${\bf P}= \frac{{\bf\bar{C}}(0)}{||{\bf \bar{C}}||}$ and should be fixed, subjected to the condition of Eq. (\ref{ts}) and is not arbitrary 
(as will be discussed during the construction of the conserved quantities) in contrast to the corresponding vector NLSE without the linear term 
characterized by the matrix ${\bf A}$. It should be noted that the solution for ${\bf Q}$ is exactly of the form ${\bf Q}= e^{-i {\bf A}t} {\bf \tilde{Q}}$ 
with ${\bf U=e^{-i {\bf A}t}}$ being a pseudo-unitary transformation as discussed in Ref. \cite{pkg}. It is easy to check that ${\bf \tilde{Q}}$ is 
the solution of Eq. (\ref{lv}) with ${\bf A}=0$, and have the following form

\bea
{\bf \tilde{Q}}= 2 \eta e^{-i\{2\xi x-4(\xi^2-\eta^2)t-\frac{\pi}{2}\}}\text{Sech}{(2\eta x-8\xi \eta t-2\delta)} {\bf P}.
\eea
However, it should be noted that ${\bf \tilde{Q}}$ represents a particular class of solutions of the vector NLSE without the linear term,
in which case the polarization vector is determined by the condition of Eq. (\ref{ts}).
Therefore, we have the important conclusion that for integrable vector NLSE with a linear term characterized, in general, by a non-hermitian matrix,
the inverse scattering transformation selects a particular class of solutions of the corresponding vector NLSE without the linear term and map it
to the solution of the integrable vector NLSE with the linear term via a pseudo unitary transformation through the relation  
${\bf Q}= e^{-i {\bf A}t} {\bf \tilde{Q}}$. 

{\bf Conserved quantities:} As has been observed in section-2, that the scattering coefficients ${\bf \bar{a}}$ and $\bf a$ are time 
independent. These coefficients yield a Lorent series expression \cite{aab} in terms of the potentials $\bf Q, R$ and their derivatives
with each term being constant. Since the Lorent series has an infinity number of terms, we get an infinite number of conserved quantities
in this case. Some of the conserved quantities may be expressed, corresponding to the series of $\bf \bar{a}$, as:

\bea
\bar {\Gamma_1}&=& - \int^{\infty}_{-\infty} {\bf RQ} dx= \int^{\infty}_{-\infty} {\bf \tilde{Q}}^{\dagger}{\bf G} {\bf \tilde{Q}}dx,\nonumber\\
\bar {\Gamma_2}&=& - \int^{\infty}_{-\infty} {\bf R}{\bf Q}_{x} dx= \int^{\infty}_{-\infty} {\bf \tilde{Q}}^{\dagger}{\bf G} {\bf \tilde{Q}_x}dx,\nonumber\\
\bar {\Gamma_3}&=&  -\int^{\infty}_{-\infty} [{\bf R}_x{\bf Q}_x - {(\bf RQ})^2]dx= \int^{\infty}_{-\infty} [{\bf \tilde{Q}_x}^{\dagger}{\bf G} {\bf \tilde{Q}_x}- 
({\bf \tilde{Q}}^{\dagger}{\bf G} {\bf \tilde{Q}})^2]dx.
\eea
It should ne noted that the conserved quantities are not positive definite in general and depend on the form of the matrix ${\bf G}$.
Similar expression of conserved quantities can be obtained for the expression of ${\bf a}$. However, it should be noted that ${\bf a}$ is a
$2\times 2$ matrix operator and is a constant provided the condition of Eq. (\ref{ts}), i.e $[{\bf A}, {\bf a}]=0$, is satisfied. The Lorentz 
series expression of ${\bf a}$, \cite{aab} suggests that this condition is satisfied provided ${\bf A}{\bf h}={\bf h}{\bf A}^{\dagger}$, 
where ${\bf h}={\bf \bar{C}}(0){\bf \bar{C}}(0)^{\dagger}$ is a singular hermitian matrix.
Therefore, the integration constant ${\bf \bar{C}}(0)$, so far appears to be arbitrary must be fixed properly. This is an important result and appears 
solely due to the inclusion of the linear term characterized by the matrix $\bf A$.

As an example, we may consider the case discussed in Ref. \cite{pkg}, and determine the form of the norming constant.
We take ${\bf A}$, ${\bf G}$ and ${\bf \bar{C}}(0)$ of the following form,

\bea
{\bf A}=\beta_1 {\bf \sigma}_1+\beta_2{\bf \sigma}_2+i \Gamma{\bf \sigma}_3, \  \  \  \ 
{\bf G}=\sum^3_{j=0} \alpha_j {\bf \sigma}_j,  \ \ \ \  {\bf \bar{C}}(0)= (\bar{C}_1 e^{i\bar{\theta}_1}, \bar{C}_2e^{i\bar{\theta}_2})^{T}
\label{agc}
\eea
where ${\bf \sigma}_j, j=1,2,3$ are the Pauli matrices and ${\bf \sigma_0}$ is the $2\times 2$ identity matrix.
The real parameters $\beta_{1,2}$ are responsible for the coupling between the complex fields and $\Gamma$
measures the loss/gain strength. The real parameters $\alpha_0, \alpha_3$ determine the self-phase and cross-
phase modulation where the real parameters $\alpha_1, \alpha_2$ are responsible for the four-wave mixing.
The matrix ${\bf A}$ is ${\bf G}$-pseudo hermitian for

\bea
\alpha_3=0,\ \ \ \  \alpha_0=\frac{|\alpha||\beta|}{\Gamma}\sin{(\theta_{\alpha}-\theta_{\beta})}
\eea
 withe the choice $|\beta|>|\Gamma|$ with $0<\theta_{\alpha}-\theta_{\beta}<\pi$ for $\Gamma> 0$ and
$\pi<\theta_{\alpha}-\theta_{\beta} <2\pi$ for $\Gamma< 0$ and $\alpha\equiv \alpha_1+i\alpha_2=|\alpha|e^{i\theta_\alpha}$,
$\beta\equiv\beta_1+i\beta_2=|\beta|e^{\theta_\beta}$. Further, the condition ${\bf A}{\bf h}={\bf h}{\bf A}^{\dagger}$ implies
$\bar{C}_1=\bar{C}_2=\bar{C}$,  $\sin{(\bar{\theta}_{12}+\theta_{\beta})}=\frac{\Gamma}{|\beta|}$, with 
$\bar{\theta}_{12}=(\bar{\theta}_{1}-\bar{\theta}_{2})$ and with the choice $|\beta|>|\Gamma|$ $0<\bar{\theta}_{12}+
\theta_{\beta}<\pi$ for $\Gamma> 0$ and $\pi<\bar{\theta}_{12}+\theta_{\beta} <2\pi$ for $\Gamma< 0$.
Therefore, finally we have the following expressions for ${\bf A}$, ${\bf G}$ and ${\bf \bar{C}}(0)$:

\bea
{\bf A}=\beta_1 {\bf \sigma}_1+\beta_2 {\bf \sigma}_2+i \Gamma{\bf \sigma}_3, \  
{\bf G}=\frac{|\alpha||\beta|}{\Gamma}\sin{(\theta_{\alpha}-\theta_{\beta})}{\bf \sigma}_0+\alpha_1{\bf \sigma}_1+\alpha_2{\bf \sigma}_2,
 \ {\bf \bar{C}}(0)= \bar{C}(e^{i\bar{\theta}_1}, e^{i\bar{\theta}_2})^{T}.
\label{cq}
\eea
Thus the norming constant and hence the polarization vector ${\bf P}$ is fixed depending on the possible form of the matrix
${\bf A}$. This is an important result and solely appears due to the inclusion of the linear term, characterized by the 
matrix $\bf A$, in case of vector NLSE.

\section{One soliton solution for the reduction ${\bf R}= -({\bf GQ})^{P}$}

In this section, the one soliton solution for the reduction ${\bf R}= -({\bf GQ})^{P}$ is presented with
the construction of the corresponding conserved quantities.
The condition ${\bf R}= -({\bf GQ})^{P}$ imposes a symmetry on the scattering data and
it may be observed from the Wronskian calculation of the scattering data that if $k_j$ is an eigen
value of the scattering data in the upper half of $k$ plane then $k^*_j$ is also  an eigen value \cite{ab,ds}. Similarly,  
if $\bar{k}_j$ is an eigen value of the scattering data  in the lower half of $k$ plane then $\bar{k}^*_j$ is 
also  an eigen value. In the subsequent discussion, without any loss of generality, we shall assume $k_j=i\eta_j$
and $\bar{k}_j=-i\bar{\eta}_j$, i.e, the eigen values are taken to be on the imaginary axis.
The one soliton solution is obtained by taking $J=\bar{J}=1$ and closing the system by putting
$k=k_1$ and $k=\bar{k}_1$ in Eq. (\ref{e6}). The expressions of 
${\bf N}$ and ${\bf \bar{N}}$, thus obtained, can now be used in Eq. (\ref{qr}) together with the time 
dependence of the norming constants as given by  Eq. (\ref{nc}) to obtain the following expressions for the 
potentials (omitting the subscript for conveniences):

\bea
{\bf Q}=2i \frac{e^{-2\bar{\eta}x} e^{-4i\bar{\eta}^{2}t} e^{-i{\bf A}t} {\bf \bar{C}}(0)
}{1+ e^{i\theta} e^{-2x(\eta+\bar{\eta})}   e^{-4i(\bar{\eta}^{2}-\eta^{2})t}}, \ \ 
{\bf R}=-2i \frac{e^{-2\eta x} e^{4i \eta^{2}t} {\bf C}(0) e^{i{\bf A}t} 
}{1+ e^{i\theta} e^{-2x(\eta+\bar{\eta})}   e^{-4i(\bar{\eta}^{2}-\eta^{2})t}}
\eea 

with the following conditions are satisfied due to the symmetry induced by ${\bf R}= -({\bf GQ})^{P}$:

\bea
|{\bf C}(0){\bf \bar{C}}(0)|=(\eta+\bar{\eta})^2,\ \ \ \ 
{\bf \bar{C}}^{\dagger}(0){\bf G}{\bf \bar{C}}(0)= (\eta+\bar{\eta})^2
\label{cond2}
\eea
and we take ${\bf C}(0){\bf \bar{C}}(0)= (\eta+\bar{\eta})^2 e^{i(\theta+\pi)}$.
In this case also, the solution for ${\bf Q}$ can be expressed in the form ${\bf Q}= e^{-{\bf A}t} {\bf \bar{Q}}$,
with ${\bf U=e^{-i {\bf A}t}}$ being a pseudo-unitary transformation. The vector
${\bf \bar{Q}}$ is the solution of the Eq. (\ref{nlv}) with ${\bf A}=0$ and have the following form:

\bea
 {\bf \bar{Q}}=2i \frac{e^{-2\bar{\eta}x} e^{-4i\bar{\eta}^{2}t} {\bf \bar{C}}(0)
}{1+e^{i\theta} e^{-2x(\eta+\bar{\eta})}   e^{-4i(\bar{\eta}^{2}-\eta^{2})t}}.
\label{gnl}
\eea
This is an important result in study of integrable non-local vector NLSE with generic non-linearity including self-phase and cross-phase
modulation with four wave mixing and it seems that the inverse scattering transformation for this model has not been considered in the
literature. The solution of Eq. (\ref{gnl}) encounters a singularity at a finite time $t=T$; at $x=0$, $T$ is given by 

\bea
T=\frac{(2n+1)\pi-\theta}{4(\eta^2-\bar{\eta}^2)}, \ \ \ n\in {\mathbb Z}.
\eea 
As in the corresponding local case the vector ${\bf \bar{C}}(0)$ is not arbitrary and should be determined by the condition of Eq. (\ref{ts}). 
Therefore, we have the important conclusion that for integrable non-local vector NLSE with a linear term characterized, in general, by a non-hermitian matrix,
the inverse scattering transformation selects a particular class of solutions of the corresponding non-local vector NLSE without the linear term and map it
to the solution of the integrable non-local vector NLSE with the linear term via a pseudo unitary transformation.

{\bf Conserved quantities:} The conserved quantities are obtained in a similar manner as in the case of the local vector NLSE
with the linear term. However, in this case the hermitian conjugation is replaced by the hermitian conjugation plus the parity
transformation of the spatial coordinates that incorporates the non-local effect. In this case also, we get an infinity number of conserved
quantities. Some of the constant quantities may be expressed, corresponding to the series of $\bf \bar{a}$, as:

\bea
\bar {\Gamma_1}&=& - \int^{\infty}_{-\infty} {\bf RQ} dx= \int^{\infty}_{-\infty} {\bf \bar{Q}}^{\bf P}{\bf G} {\bf \bar{Q}}dx,\nonumber\\
\bar {\Gamma_2}&=&  -\int^{\infty}_{-\infty} {\bf R}{\bf Q}_{x} dx= \int^{\infty}_{-\infty} {\bf \bar{Q}}^{\bf P}{\bf G} {\bf \bar{Q}_x}dx,\nonumber\\
\bar {\Gamma_3}&=& - \int^{\infty}_{-\infty} [{\bf R}_x{\bf Q}_x - {(\bf RQ})^2]dx= \int^{\infty}_{-\infty} [{\bf \bar{Q}_x}^{\bf P}{\bf G} {\bf \bar{Q}_x}- 
({\bf \bar{Q}}^{\bf P}{\bf G} {\bf \bar{Q}})^2]dx.
\eea
It is interesting to note that none of the above conserved quantities are hermitian even for a positive definite ${\bf G}$. 
This is due to the non-local nature of the reduction
of Eq. (\ref{re2}). However, some of the conserved quantities can be shown to be real valued by splitting the fields as a sum of parity even and
parity odd terms \cite{ds,ds1}.  Similar expression of conserved quantities can be obtained for the expression of ${\bf a}$. 
However, as in the corresponding local case, the integration constant ${\bf \bar{C}}(0)$ should be fixed subjected to the condition
of Eq. (\ref{ts}), i.e $[{\bf A}, {\bf a}]=0$. The Lorentz series expression of ${\bf a}$, suggests that this condition is satisfied provided 
${\bf A}\{{\bf \bar{C}}(0){\bf \bar{C}}(0)^{\dagger}\}
=\{{\bf \bar{C}}(0){\bf \bar{C}}(0)^{\dagger}\}{\bf A}^{\dagger}$, where ${\bf \bar{C}}(0){\bf \bar{C}}(0)^{\dagger}$ is a singular hermitian matrix.
Therefore, the integration constant ${\bf \bar{C}}(0)$, so far appears to be arbitrary must be fixed properly. 
As an example, we may chose the possible form of ${\bf A}$, ${\bf G}$ and ${\bf \bar{C}}(0)$ as given by Eq. (\ref{agc}). The same
calculation as in the corresponding local case yields Eq. (\ref{cq}), however, in this case the condition (\ref{cond2}) further specifies the results
with $\bar{C}^2=\frac{(\eta+\bar{\eta})^2}{2[\alpha_0+|\alpha|\cos{(\bar{\theta}_{12}+\theta_{\alpha})}]}$, with the choice $\alpha_0>|\alpha|$. 
Thus the norming constant and hence the polarization of the solution is fixed depending on the possible form of the matrix
${\bf A}$. This is an important result and solely arises due to the inclusion of the linear term, characterized by the 
matrix $\bf A$, in the case of non-local vector NLSE.

\section{Summary and discussions}

We have studied the integrable properties of local and non-local vector NLSE with generic form
of cubic non-linearity in presence of a linear term characterized, in general, by a non-hermitian matrix
which under certain condition incorporates balanced loss and gain with a linear coupling between the 
complex fields of the vector NLSE. The cubic interaction includes self-phase and cross phase modulation 
along with four wave mixing. The non-local model considered may be viewed as a two fold generalization 
of non-local vector NLSE considered earlier in the literature \cite{ds}.  In particular, it generalizes the cubic
interaction to include the self-phase and cross-phase modulation along with four wave mixing. Further, the
integrability of this model has been investigated in presence of a linear term characterized, in general, by
a non-hermitian matrix which under particular situation corresponds to balanced loss and gain with a linear 
coupling between the complex fields of the non-local vector NLSE. An important corollary of the present 
investigation is the study of the integrable properties of non-local vector NLSE, only with generic cubic 
non-linearity involving the self-phase and cross phase modulation along with four wave mixing and obtaining 
the exact soliton solution by using inverse scattering method. It seems that the inverse scattering transformation 
for non-local vector NLSE model with the generic form of the cubic non-linearity has not been considered earlier in 
the literature. The Lax-pair for vector NLSE with the linear term have been constructed for both the local and non-local 
reductions and an infinite number of conserved quantities are found. This confirms the integrability of the systems. 
Apart from the particular form of the reduction, the systems are shown to be integrable only when the matrix representing
the linear term is pseudo hermitian with respect to the matrix comprising the generic cubic non-linearity.
This is an important result in the realm of integrability of NLSE with a linear term and having a generic cubic non-linearity. The inverse 
scattering transformation method has been employed to find exact soliton solutions for the vector NLSE with the linear term 
for the local as well as non-local cases. Another important observation of the present investigation is that
the presence of the linear term restricts the possible form of the norming constants and hence the polarization vector. 
It is shown that for integrable vector NLSE with a linear term characterized, by a pseudo-hermitian matrix, the inverse
scattering transformation selects a particular class of solutions of the corresponding vector NLSE without the linear term and 
map it to the solution of the integrable vector NLSE with the linear term via a pseudo unitary transformation, for both the local and non-local
cases. It should be mentioned that although some exactly solvable models are constructed in the context of NLSE incorporating
balanced loss and gain and linear coupling, the integrable properties of such systems are not investigated previously in the literature. 
Further, two integrable systems are said to be gauge equivalent if the corresponding Lax-pairs are 
related via a gauge transformation. For the present case this gauge transformation to the corresponding integrable vector NLSE 
without the linear term is not apparent. Within this background,  the present study on the integrable properties of the local and
non-local vector NLSE in presence of a linear term characterized by a non-hermitian matrix which under certain condition incorporates 
balanced loss and gain with linear coupling between the complex fields of the vector NLSE, is an important addition to 
the existing list of integrable systems. Further, it is observed that the presence of the linear term characterized by the non-hermitian matrix
determines the possible form of the polarization vector. Therefore, the study of the multi-soliton solution and the collision dynamics 
between the solitons for the local and non-local vector NLSE with the linear term characterized, in general, by a non-hermitian matrix will be very
much interesting \cite{ds3}.

\section{Acknowledgements}

DS acknowledges a research fellowship from CSIR (ACK No: 362103/2k19/1, File No: 09/489(0125)\\/2020-EMR-I).


\begin{thebibliography}{10}

\bibitem{ac}V.E Zakharov and A.B. Shabat, Sov.Phys. JETP {\bf 34}, 62 (1972).


\bibitem{r2} L. P. Pitaevskii and S. Strinari, Bose-Einstein Condensation, Oxford
University Press, Oxford, 2003; Emergent Nonlinear Phenomena in Bose-Einstein
Condensation: Theory and Experiment, edited by P. G. Kevrekidis, D. J. Frantzeskakis
and R. Carretero-Gonz$a^{\prime}$lez(Springer, New York, 2008), Vol. 45.
\bibitem{r3}   Koji. Mio et al, 
J. Phys. Soc. Jpn. 41,  265 (1976)

\bibitem{r4}   Karsten Trulsen, Kristion B. Dysthe. 
Wave motion 24,  281 (1996).

\bibitem{r5}  S. Yomosa. Phys. Rev. A {\bf 27}, 2120 (1983).


\bibitem{cl}H. H. Chen and C. S. Liu, Phys. Rev. Lett. {\bf37}, 693 (1976);
Phys. Fluids 21, 377 (1978).

\bibitem{vn}V.N. Serkin, A.Hasegawa, T.L. Belyaeva, Phys. Rev. Lett {\bf 98},
 074102 (2007).

\bibitem{r6} Jun-Rong He and Hua-Mei Li. Phys. Rev. E {\bf 83}, 066607 (2011).


\bibitem{r8} Juan Belmonte-Beitia, Victor M. Perez-Garcia, Valerity Brazhnyi, Commun Nonlinear Sci Numer Simulat 16, 158 (1996).

\bibitem{r9}  Juan Belmonte-Beitia, Gabriel F. Calvo, Physics Letters A 373,  448 (2009).


\bibitem{ag} A. Hasegawa and Y. Kodama, Solitons in Optical Communications (Oxford University Press, New York, 1995);
L. F. Mollenauer and J. P. Gordon, Solitons in Optical
Fibers (Academic Press, Boston, 2006); G. P. Agrawal, Nonlinear Fiber Optics (Academic Press,
San Diego, 2001), 3rd ed.

\bibitem{r1}   K. G. Makris et  al, Int J Theor Phys 50. 1019 (2011).


\bibitem{r10}
 Z.H. Musslimani,  K. G. Makris, R. El-Ganainy, and D. N. Christodoulides, Phys. Rev. Lett. {\bf 100}, 030402 (2008); 
M. Duanmu, K. Li, R. L. Horne, P. G. Kevrekidis, and
N. Whitaker, Philos. Trans. R. Soc. A {\bf 371}, 20120171 (2013);
M.-A. Miri, A. B. Aceves, T. Kottos, V. Kovanis, and D. N.
Christodoulides, Phys. Rev. A {\bf 86}, 033801 (2012).



\bibitem{su} Sumei Hu, Xuekai Ma, Daquan Lu, Zhenjun Yang, Yizhou Zheng, Wei Hu, Phys.
Rev. A {\bf 84} (2011) 043818.

\bibitem{zh} Zhiwei Shi, Xiujuan Jiang, Xing Zhu, Huagang Li, Phys. Rev. A {\bf 84} (2011) 053855.


\bibitem{zh1} Z.H. Musslimani, K.G. Makris, R. El-Ganainy, D.N. Christodoulides, Phys. Rev.
Lett. {\bf 100} (2008) 030402.


\bibitem{ab} M. J. Ablowitz, Z. H. Musslimani, Phys. Rev. Lett {\bf 110}, 064105 (2013).


\bibitem{ds} D. Sinha and P. K. Ghosh, Phys. Lett. A {\bf381}, 124 (2017).


\bibitem{ds2} D. Sinha and P. K. Ghosh, Phys. Rev. E {\bf 91},
042908(2015)

\bibitem{abzm} M. J. Ablowitz and Z. H. Musslimani, Nonlinearity {\bf29}, 915(2016).

\bibitem{asf} A. S. Fokas, Nonlinearity {\bf29}, 319(2016).


\bibitem{aks} A.K. Sarma et al, Phys. Rev. {\bf E89}, 052918 (2014). 

\bibitem{akas} A. Khare, A. Saxena, J. Math. Phys. {\bf 56}, 032104 (2015).

\bibitem{akmu} A.K. Sarma, Z. H. Musslimani, Phys. Rev. {\bf E90}, 032912 (2014). 

\bibitem{li} M. Li, T. Xu, Phys. Rev. {\bf E91}, 033202 (2015). 


\bibitem{lym} Li-Yuan Ma, Zuo-Nong Zhu, J. Math. Phys. {\bf 57}, 083507 (2016)

\bibitem{xi} Xiang Li, Xiao-Tao Xie, Phys. Rev. A {\bf90} (2014) 033804.

\bibitem{igor} I.V. Barashenkov, Sergey V. Suchkov, Andrey A. Sukhorukov, Sergey V. Dmitriev,
Yuri S. Kivshar, Phys. Rev. A {\bf86} (2012) 053809.

\bibitem{yu} Yu.V. Bludov, R. Driben, V.V. Konotop, B.A. Malomed, J. Opt. {\bf 15} (2013) 064010.

\bibitem{ro} Rodislav Driben, Boris A. Malomed, Opt. Lett. {\bf 36}, 4323 (2011).

\bibitem{alj} Alejandro J. Martínez, Mario I. Molina, Sergei K. Turitsyn, Yuri S. Kivshar, Phys.
Rev. A {\bf 91} (2015) 023822.


\bibitem{pkg} P. K. Ghosh, Phys. Lett. A {\bf 402}, 127361 (2021).

\bibitem{sg} S. Ghosh and P. K. Ghosh, arXiv:2112.04802.




\bibitem{rr10} Yu. V. Bludov, V. V. Konotop, and B. A. Malomed,  Phys. Rev. A  {\bf 87}, 013816 (2013).

\bibitem{rw} C. Kharif, E. Pelinovsky, and A. Slunyaev, Rogue waves
in the ocean (Springer, Heidelberg, 2009).

\bibitem{ex1} R. Driben, and B. A. Malomed,  EPL {\bf96}, 51001 (2011).

\bibitem{ex2} V. V. Konotop, J. Yang, and D. A. Zezyulin, Rev. Mod. Phys. {\bf 88}, 035002 (2016).

\bibitem{igor1} N. V. Alexeeva, I. V. Barashenkov, Andrey A. Sukhorukov, and Yuri S. Kivshar,  Phys. Rev. A  {\bf 85}, 063837 (2012).

\bibitem{wang} D.-S. Wang, D. -J. Zhang and J. Yang, J. Math. Phys. {\bf51}, 023510(2010).

\bibitem{rl} R. Radhakrishnan, M. Lakshmanan, and J. Hietarinta, Phys. Rev. E {\bf56}, 2213 (1997).

\bibitem{tt} T. Tsuchida, Prog. Theor. Phys. {\bf111}, 151 (2004).

\bibitem{sc} V. E. Zakharov and E. I. Schulman, Physica D {\bf4}, 270 (1982).

\bibitem{av} A. V. Mikhailov, Physica D {\bf 3}, 73 (1981).

\bibitem{bl} Z. Shi and J. Yang, Phys. Rev. E {\bf75}, 056602 (2007).

\bibitem{becs} F. Dalfovo, S. Giorgini, L. P. Pitaevskii, and S. Stringari, Rev. Mod. Phys. {\bf71}, 463 (1999).

\bibitem{becs1} J. Ieda, T. Miyakawa, and M. Wadati, J. Phys. Soc. Jpn. {\bf73}, 2996 (2004).

\bibitem{ds1} D. Sinha, Under preparation.

\bibitem{pseh} A. Mostafazadeh, J. Math. Phys. {\bf45}, 932 (2004).

\bibitem{aab} M. J. Ablowitz, B.Prinari and A. D. Trubatch, Discrete and Continuous Non-linear
Schrodinger Systems.(Cambridge University Press, Cambridge, Enaland, 2004).

\bibitem{ds3} D. Sinha, Under preparation.






\end{thebibliography}
\end{document}